\documentclass[twocolumn,english]{IEEEtran}
\usepackage[T1]{fontenc}
\usepackage{array}

\makeatletter

\providecommand{\LyX}{L\kern-.1667em\lower.25em\hbox{Y}\kern-.125emX\@}
\providecommand{\tabularnewline}{\\}

\newtheorem{conject}{Conjecture}
\newtheorem{definitn}{Definition}

\makeatother

\usepackage{babel}

\begin{document}

\title{Knowledge Dispersion Index for Measuring Intellectual Capital}

\author{Vikram Dhillon%
\thanks{Electronic address: dhillonv10@gmail.com%
}}
\maketitle
\begin{abstract}
In this paper we propose a novel index to quantify and measure the
flow of information on macro and micro scales. We discuss the implications
of this index for knowledge management fields and also as intellectual
capital that can thus be utilized by entrepreneurs. We explore different
function and human oriented metrics that can be used at micro-scales
to process the flow of information. We present a table of about 23
metrics, such as change in IT inventory and percentage of employees
with advanced degrees, that can be used at micro scales to wholly
quantify knowledge dispersion as intellectual capital. At macro scales
we split the economy in an industrial and consumer sector where the
flow of information in each determines how fast an economy is going
to grow and how overall an economy will perform given the aggregate
demand. Lastly, we propose a model for knowledge dispersion based
on graph theory and show how corrections in the flow become self-evident.
Through the principals of flow conservation and capacity constrains
we also speculate how this flow might seeks some equilibrium and exhibit
self-correction codes. This proposed model allows us to account for
perturbations in form of local noise, evolution of networks, provide
robustness against local damage from lower nodes, and help determine
the underlying classification into network super-families. \end{abstract}

\section{Introduction}

Lord Kelvin is often quoted for saying that if one can't measure something
they can't improve it. This is critical in the field of knowledge
management as metric provide convincing data for entrepreneurs to
make decisions and help capitalize it for economic and social progress.
The aspiration behind developing this index arose in response to lack
of a quantifying measure for the following: 
\begin{itemize}
\item The exchange of tactical knowledge concerning one particular field
between employees. 
\item Capture of that information in a knowledge-base for use by other employees.
\item The reuse rate of frequently accessed knowledge. 
\end{itemize}
This index will be divided into two levels, each further consisting
of its own metrics, the outputs from each level can be used as independent
indicators and the gross measure will complete the Knowledge Dispersion
Index (KDI) for a given location. It must be noted that this analysis
involves human input therefore some aspects can not be captured effectively,
although we do include an assumption called the Theory Y of management
which asserts that management assumes employees may be ambitious,
self-motivated, exercise self-control and if allocated adequate resources,
they will perform at their very best. Given that assumption, we establish
the two levels of our index:
\begin{itemize}
\item Organizational knowledge metric: This metric will account for aggregate
dispersion at micro scales and indicate the growth of the industrial
sector. 
\item General dispersion metric: This metric will account for dispersion
at macro scales and indicate how the consumer side is performing.
\end{itemize}

\section{Micro Scale (Organizational Knowledge Metric)}

At the micro scale industries play a prominent role, the reason for
including industries at the micro scale pertains to their role in
the economy. The distributer's side is generally smaller in comparison
to the consumer side. In addition to that, there are some particular
mechanisms that each of the firms employ to enhance their success
and if they are not fine-tuned, the results can be disastrous. For
that very reason we present a comprehensible list (see Appendix A
for the list) to account for the flow of information, the gross result
obtained from the table in the Appendix will present a reliable measure
of the information infrastructure of a company%
\footnote{In this paper we don't provide definite benchmarks which can then
be used to compare, a case study however, regarding those benchmarks
is underway. %
}. There is however another very important role that the organizational
knowledge metric plays, that of identifying the most efficient method
of dispersion and the faults that may occur in the dispersion through
the use of a theoricial construction called a flow networks.

\subsection{Flow Networks}

Flow networks are the key structure in network analysis and here we
show how the traditional flow network model can be employed by a company
to analyze the flow of information in their infrastructure. First
we review the traditional flow network model \cite{key-1}, a flow
network is a directed graph where each edge has a capacity and each
edge receives a flow. The amount of flow on an edge cannot exceed
the capacity of the edge. A flow must satisfy the restriction that
the amount of flow into a node equals the amount of flow out of it,
except when it is a source, which has more outgoing flow, or sink,
which has more incoming flow. 

Formally, $G(V,\, E)$ is a directed graph where each edge $(u$$\, v)\:\epsilon\, E$
has a non-negative, real valued capacity given by a function $c(u\, v)$.
There are however two special vertices where the capacity constrains
don't hold: a source $s$ and a sink $t$. The rest of the nodes $u,\, v$
follow the following rules: 
\begin{itemize}
\item Capacity Limit: $f\,(u,\, v)\:\leq c\,(u,\, v).$ This relation implies
that the flow to one node can not exceed the capacity limit given
to that node. 
\item Skew Symmetry: $f\,(u,\, v)\;=-f\,(u,\, v).$ This relation implies
that the net flow from one node to another or from $u$$\rightarrow v$
must be the same from $v\rightarrow u$. 
\item Flow conservation: $\sum\, f\,(u,\, w)=k.$ This relation can be seen
from the two above, the net flow to a node is some constant $k$ which
holds the same value for each node at the same hierarchy but changes
value at different levels of management, except for the two cases
of source and sink. 
\end{itemize}
Now we will explore how a model following these rules can help us
explore the efficiency of the information infrastructure. First we
see that the capacity limit holds to how much information should be
made avialable to employees at each stage, this limit helps us understand
better where in the graph is too much information is present as that
in itself is a loss or waste since one employee can only utilize so
much of what is given to them. Knowing the capacity limit for the
nodes can also help us by reducing the flow to one stressed node by
spreading it amongst other available nodes at the same level of hierarchy.
The skew symmetry serves as a test that can be employeed by the management
to asses whether each employee at their current level is utilizing
their resources to their maximum or not, the full employement of the
resources at each level would result in for instance completion of
more patents being filed or more solutions being completed by a company
as the demand for that particular solution increases. It must be noted
here that this relation holds dual values, the flow of information
decreases as it reaches the top as can be seen in the case of a software
development company, the CEO of the company is generally well-versed
in business techniques and not in lower-level coding therefore the
higher-level overview being presented to them in the form of a flowchart
would be represented by a decrease in the flow of information reaching
them. This should not be considered a short-coming of the model because
each level of hierarchy will have a constant flow, so as information
moves up that hierarchy and reached the management positions of a
company, the flow will decrease and stabalize. The third rule indicates
the communication infrastructure present in a company occuring among
various nodes (employees) at each level. A company generally has several
different departments or smaller level committes to perform various
tasks, this rule applies to how well each of those departments interact
with the other ones to finish a given task. The other two rules provided
before can also be applied here to each of the departments to obtain
more specifics. We argue that modelling a company as a flow model
and following a combination of these three rules will provide a very
good estimate of the information flow in a company. The data requried
to asses these rules will be provided by the gross result from the
list in Appendix A.

\subsubsection{Self Correction Codes}

Using the flow model for information dispersion also gives us the
ability to use the inherent correction features present in the flow
model. We assert that a company modeled as a traditional flow model
will have the tendency to correct itself as needed, this is also a
basic assumption in game theory, that if each employee is seen a player
in a game, each player will make the best possible move. We generalize
that assertion in the following:
\begin{conject}
If leadership is emplyoed in a multi-player show definitegame, the
leader will make the best possible choice for the players.
\end{conject}
This will help in the evolution of the network, as the company prospers
the network will expand and stabalize for definite capacity limits
at every level, this is true of a successful company. For a newly
developing company, the governing dynamics will be different, each
new employee is in a learning stage but are required to follow skew
symmetry and flow conservation. As a result of that, there will be
a discrepancy in the rules as the new employee will be a pseudo-node
and this discrepancy can be analyzed in terms of a pertubation to
the model. The flow model approach provides a solution for the model
to be robust against such changes, to address local pertubations,
we propose the following: 
\begin{conject}
Given a recently established company, and a measure of reliability
determined from our Appendix, some nodes with a high measure of reliability
should be overloaded. 
\end{conject}
It should be noted here that overloading those nodes doesn't necessarily
imply a decrease in efficiency of those employees, the idea is to
vary the flow amongst those reliable nodes and split flow conservation
into two levels, one for the newer employees and the other for the
reliable ones. This allows the learning stage of the new employees
which acts as a pertubation to be absorbed in by the overloading of
those nodes. The output provided by the newer pseudo-nodes %
\footnote{The new employees only act as pseudo-nodes while they are under some
initial training, as soon as they are ready, we decrease the flow
to the nodes at which the new employees are to work, that allows a
smooth introduction of new nodes in the flow.%
} will also go through a similar process but the difference is that
the output gets absorbed by the these overloaded nodes.

Now we will address how network super-families arise in this evolved
flow, from Conjecture 2.2 we determined a reliability measure that
can help use overload some nodes, those nodes at each hierarchical
level will constitute our network super-family. Formally, the members
of the yet to be formed superfamiliy will appear as outliers if a
statistical analysis such as ANOVA is performed because of their reliability
profile established from the Appendix metrics. Once super-families
are identified in a flow, our network becomes very strong and is robust
to local and internal damage. 
\begin{conject}
Given a flow network with identified network super-families, local
damage can be diverted to specified nodes, decreasing its impact. 
\end{conject}
Local or internal damage lacks a formal operational definition so
we have to rely on a situational definition. Any intentional efforts
to disrupt a well established flow of information constitutes local
damage, if the causes of which are identified, for instance an attack
targeting some confidential information can be diverted to the pseudo-nodes,
rendering it useless since the pseudo-nodes have lower level clearance.

\section{Macro Scale (General Dispersion Metric)}

We now analyze how information disperses amongst the general public
or the consumer side. It must be noted here that KDI would only be
compelte as an index once we can get results from both the metrics,
implying that we need to quantify how information flows amongst industries
which form the micro scale and the consumer side or the general public
which form the macro scale. The methods of analysis being employed
before simply become inefficient here because of the graphes modelling
a large amount people would easily break down and is quite inefficient
computationally. There are however some metrics that can help us here
such as: 
\begin{itemize}
\item Transport infrastructure
\item Availability of communications facilities such as television and internet
\item Public spending on broadcasting
\item Access to internet
\item Frequency of local awareness campaigns such as seminars and so on.
\end{itemize}
We will here employ another theoretical construction to analyze how
information is spread to the general population or the consumers sector.
Some charateristics of this study are that it must be general enough
to account for a vast variety of sources and at the same time be able
to provide a definite analysis for the comparison of the output from
each source.
\begin{definitn}
There are several sources that aid the process, such as television,
newspaper, etc. however, for simplicity of analysis, we will collectively
refer to these different sources as media. 
\end{definitn}

\subsection{Deep Current Sets}

The most common source for information available to the public%
\footnote{We simply refer it to as a source, its arguable which source constitues
as the most widely available for different regions.%
} outputs vast amounts of interpreted information, this output can
be viewed as a surface wave or a current in an ocean. This analogy
of ocean currents is being used because the currents themselves are
very dynamic and based on very delicate balance, we imply that in
the same degree as ocean currents on the surface can influence the
weather pattern of the atmosphere above, the general croud is influenced
by the larger waves or the media source that is most influencial,
imposing similar constant weather conditions over a large portion
of the atmosphere. This construction helps analyze the behaviour and
choices of the people in the given region in the past to make deductions
about the future, from that we determine how the information flow
that once only occured amongst a small group of people has evolved
now to spread to much larger networks. We present a worked example
of deep current sets in the light of a hypothetical developing situation.
\begin{conject}
Given an area and a developing situation, the flow of information
is given as follows: 

Before the development: Those involved with an event may share or
record their ideas, theories, or plans alone in a lab or personal
journal, with friends or colleagues. The sharing of this information
forms the \textquotedbl{}invisible college.\textquotedbl{} Usually,
the ideas generated by the \textquotedbl{}invisible college\textquotedbl{}
are simply not available to the public and is nonexistent as a form
of information today, especially for things in the distant past. This
sets the stage for the currents to arise in the ocean. 

As the development occurs: Early news releases may appear on TV or
radio, in newspapers, over newswires, and on the Web. The initial
information about these events generally accompnies the \textquotedbl{}who\textquotedbl{},
\textquotedbl{}what\textquotedbl{}, \textquotedbl{}where\textquotedbl{}
and \textquotedbl{}when\textquotedbl{} and most often exclude the
\textquotedbl{}why\textquotedbl{} 

The next day or the days after: Articles appear in newspapers/newswires;
information is disseminated on TV, radio, and the Web. Depending on
the event, the information may be prolific or sparse. 

In a week or the weeks after: Articles may appear in general or subject-focused
popular magazines. 

In a month or time that follows: Articles appear in scholarly journals.
This is also when scholars and researchers may start holding conferences
on the topic and eventually, conference papers will be published. 
\end{conject}
We see how through the development of this situation the ocean currents
are forming, the before sets up the currents and the next phase is
when the currents are out in the ocean and their relative size determines
how the weather is changes. From the changes in weather which are
easily observable, we can reliably make deduction about the past history
of the people. The advantage of this analysis provides is that even
though a given region may have changed now, we can still see the effects
of the currents in the future%
\footnote{This in some sense is similar to a pebble skipping across water, whereever
it lands, even for a brief moment of time, it leaves behind ripples
that can be the analyzed to trace its path.%
}, this analysis of the past can also assist in the development of
better dynamics modelling the behaviour of that region. In addition
to that, we can also determine what forms of media are most prevalent
in that given region. This analysis can be conducted based on a past
event and hold its validity for a long time. 

As we see, the quantization of information flow here can be further
assisted through the use of metrics such as \% access to cable, people
with advanced degrees, average household income and so on. This can
help determine the interests of people and classify them into sets,
however in our construction, we can construct computer models based
on using restricted boltzmann machines through a technique called
deep learning. If a technology firm sees an opportunity to start developing
in a given region, if they can identify the type of issues being discussed
amongst the community (where people can be organized into different
sets based on interests), they can identify the region%
\footnote{Formally, where the largest set is interested in the same objective
as the company%
} where it is best for them to work and since people in that community
are willing to work on those issues, establishing new businesses there
to give them an opportuinity is the best way to unleash intellectual
power.

\section{Conclusion}

In this paper, we presented how intellectual capital can be measured
through the use of a novel index, KDI. We also discussed how the metrics
present in the index can help increase the efficiency of an organization
by modeling the structure of a company using network flow models.
The last section discussed how the information flow among people in
a given region can help organizations unleash intellectual power and
capitalize on it by providing the people the necessary resources.
We are currently unable to provide numerical analysis of our index
and therefore we are lacking milestones that can provide comparative
analysis however a case study to quantify those milestones is being
undertaken. Further research would include the aforementioned milestones
and a more sophisticated models of flow networks. 

I would like to thank Christopher Norris for helpful discussions in
the writing of this paper. 

\appendix

\section{Appendix }

Here is the list of metrics \cite{ahmed02} that we will be using
at micro scales (mostly applicable to companies/firms), we do make
two assumptions here: 
\begin{itemize}
\item These metrics can be easily obtained from available data and this
will become self-evident to some extent.
\item The use of IT and various related means yields to the best dispersion
of knowledge/information within an organization.
\end{itemize}
\begin{tabular}{|l|}
\hline 
Patents pending\tabularnewline
\hline 
\% Investment in IT\tabularnewline
\hline 
\% R\&D invested in basic research\tabularnewline
\hline 
Average years of service with the company\tabularnewline
\hline 
Number of employees\tabularnewline
\hline 
Number of managers\tabularnewline
\hline 
Average duration of employment\tabularnewline
\hline 
Number of new solutions/products suggested\tabularnewline
\hline 
New patents/software/etc. filed\tabularnewline
\hline 
IT development expense/IT expense\tabularnewline
\hline 
Average age of employees\tabularnewline
\hline 
IT literacy of a staff\tabularnewline
\hline 
Company managers with advanced degrees\tabularnewline
\hline 
Revenues resulting from new business operations\tabularnewline
\hline 
IT performance/employee\tabularnewline
\hline 
IT capacity (CPU and DASD)\tabularnewline
\hline 
Changes in IT inventory\tabularnewline
\hline
\end{tabular}

\end{document}